\documentclass[aip,preprint
 aip,
 amsmath,amssymb,
 reprint,%
]{revtex4-1}

\usepackage{graphicx}
\usepackage{dcolumn}
\usepackage{bm}

\usepackage[utf8]{inputenc}
\usepackage[T1]{fontenc}
\usepackage{mathptmx}
\usepackage{etoolbox}

\usepackage{xcolor}
\usepackage{natbib}

\makeatletter
\def\@email#1#2{%
 \endgroup
 \patchcmd{\titleblock@produce}
  {\frontmatter@RRAPformat}
  {\frontmatter@RRAPformat{\produce@RRAP{*#1\href{mailto:#2}{#2}}}\frontmatter@RRAPformat}
  {}{}
}%
\makeatother
\begin{document}

\preprint{AIP/123-QED}

\title{Bundling instability of lophotrichous bacteria}
\author{Jeungeun Park}
 \affiliation{Department of Mathematics, State University of New York at New Paltz, NY 12561, USA}
\author{Yongsam Kim}%
\affiliation{ Chung-Ang University, Seoul 06974, Republic of Korea
}%

\author{Wanho Lee}
\affiliation{%
National Institute for Mathematical Sciences, Daejeon 34047, Republic of Korea
}%

\author{Veronika Pfeifer}
\author{Valeriia Muraveva}
\affiliation{Institute of Physics and Astronomy, University of Potsdam, 14476 Potsdam, Germany}

\author{Carsten Beta}
\affiliation{Institute of Physics and Astronomy, University of Potsdam, 14476 Potsdam, Germany}
\affiliation{Nano Life Science Institute (WPI-NanoLSI), Kanazawa University, Kanazawa 920-1192, Japan}

\author{Sookkyung Lim}
\altaffiliation{Author to whom correspondence should be addressed: sookkyung.lim@uc.edu}
\affiliation{Department of Mathematical Sciences, University of Cincinnati, Cincinnati, OH 45221, USA}

\date{\today}

\begin{abstract}
We present a mathematical model of lophotrichous bacteria, motivated by {\it Pseudomonas putida}, which swim through fluid by rotating a cluster of multiple flagella extended from near one pole of the cell body. Although the flagella rotate individually, they are typically bundled together, enabling the bacterium to exhibit three primary modes of motility: push, pull, and wrapping. One key determinant 
of these modes is the coordination between motor torque and rotational direction of motors.  The computational variations in this coordination
reveal a wide spectrum of dynamical motion regimes, which are modulated by hydrodynamic interactions between flagellar filaments.
These dynamic modes can be categorized into two groups based on the collective behavior of flagella, i.e., bundled and unbundled configurations.
For some of these configurations, experimental examples from fluorescence microscopy recordings of swimming {\it P.~putida} cells are also presented.
Furthermore, we analyze the characteristics of stable bundles, such as push and pull, and investigate the dependence of swimming behaviors on the elastic properties of the flagella.
\end{abstract}

\maketitle

\section{\label{sec:Introduction}Introduction}

Flagellated bacteria exhibit distinct flagellar arrangements across species, resulting in diverse locomotion strategies as they swim \cite{grognot2021more,thormann2022wrapped}. 
For instance, monotrichous bacteria such as {\em Vibrio alginolyticus} exhibit a run-reverse-flick motion\cite{ Homma1996, Magariyama2001,Son2013, Xie2010}, 
while peritrichous bacteria such as {\em Escherichia coli}
display a run-and-tumble motion, allowing movement in various directions and enhancing their capacity to explore diverse environments\cite{Berg2003, Darnton2007, turner2016visualizing}. Amphitrichous bacteria such as {\em Magnetospirillum magneticum} possess two flagella at each pole and coordinate flagellar rotation, which exhibits three modes of motility: runs, tumbles, and reversals. Asymmetric and symmetric rotations of flagella lead the cell to run and tumble, respectively\cite{Murat2015}.

Lophotrichous bacteria under investigation in this study, exemplified by {\it Pseudomonas putida}, are
characterized by the presence of a tuft of left-handed helical flagella attached near one pole of the cell body.
This distinctive flagellation pattern commonly demonstrates pull-wrapping-push motion, incorporating occasional pauses of motor rotation.\cite{harwood1989flagellation,alirezaeizanjani2020chemotaxis,hintsche2017polar,pfeifer2022role}
Push and pull modes are directed movements where the cell moves forward and backward, respectively.
When all motors rotate synchronously either counterclockwise (CCW) or clockwise (CW), a cohesive flagellar bundle is formed and
propels the cell body either forward by pushing or backward by pulling.
While motors turning CW, an increase in motor torques triggers buckling of the flagellar bundle, which then wraps around the cell body\cite{hintsche2017polar, park2022modeling}. In this case, the cell swims straight in a corkscrewed manner, allowing it to
change swimming direction, followed by the push mode.
It is known that these smooth run episodes require the ability to form a bundle which critically depends on synchronous operation of the flagellar motors.
The formation of a coherent bundle is crucial in
bacterial chemotaxis as it facilitates smooth and efficient swimming in the desired direction\cite{qu2018changes}. 

Flagellar bundling, characterized by the coordinated synchronization of motor rotations, has been extensively studied, primarily within peritrichous bacteria {\it E. coli} (see the literature\cite{lee2023bio, kim2003macroscopic, kim2004particle,lee2018bacterial, lim2012fluid, powers2002role,
macnab1977bacterial} and the references therein). Many factors, such as flagellar geometry, flagellar arrangement, counterrotation of the cell body, and 
hydrodynamic interactions between the flagellar filaments, have been identified as contributors to the formation of stable bundles. 
However, the investigation of flagellar bundling of lophotrichous bacteria remain ongoing, with a particular interest in elucidating
the mechanisms underlying flagellar interactions and the synchronization of flagellar rotations\cite{grognot2021more,nguyen2018impacts}.

In our previous work\cite{park2022modeling}, we built a mathematical model of a lophotrichous bacterium with a single flagellar bundle, which is represented as a single filamentous structure, and reproduced a typical sequence of swimming modes, pull-wrapping-push motion. Furthermore, we investigated cell reorientation in three dimensions and found that transitions from wrapping to push modes, with intermittent pauses, determine new trajectories, and the reorientation direction depends on pause timing and duration.
In this work, we extend our previous model of a swimming lophotrichous bacterium {\it P. putida} in a fluid. This model organism under consideration consists of a
spherocylindrical cell body and two identical flagella for simplicity. These flagella are affixed perpendicular to the cell surface in close proximity
to one pole of the cell body, exhibiting an equal angle relative to the body axis.
It is obvious that more flagella can be attached as needed. Our investigation focuses on the bundling instability of flagellar dynamics under various conditions of 
applied torque, direction of motor rotation, and elastic property of flagella. Additionally, we scrutinize stable bundles, with a particular emphasis on 
pull and push modes.
Finally, while the push, pull, and wrapping modes are well known from earlier literature\cite{hintsche2017polar,alirezaeizanjani2020chemotaxis, park2022modeling}, we present experimental evidence for some of the less common swimming modes that are predicted by our numerical simulations.

\section{\label{sec:model}Mathematical model}

\begin{figure}[tp!]
	\centering
\includegraphics[width=1.0\linewidth]{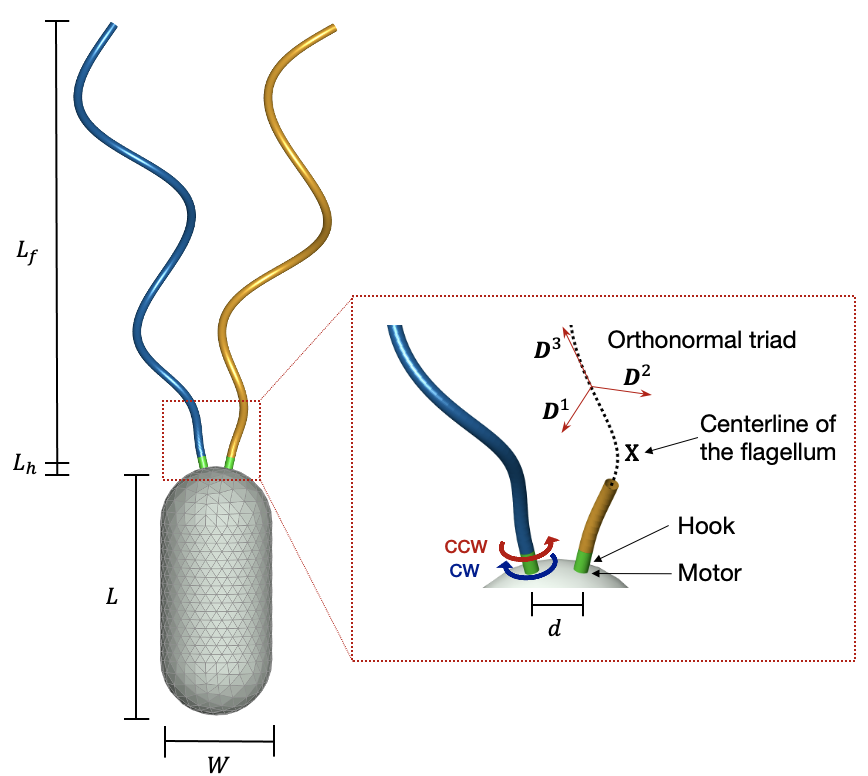}
	\caption{
A schematic diagram of our computational model.  Sketch of the initial shape of the cell body and flagella in the default setting. The inset describes details of the computational model of the flagella.}
 \label{figure}
\end{figure}

Our mathematical model is constructed based on the geometric characteristics of a microorganism {\it P. putida},
which is a lophotrichous bacterium that has a rod-shaped cell body and
a tuft of flagella attached near one pole of the cell body. This model organism is immersed in a viscous fluid and propels itself through the fluid by rotating flagella.
Although {\it P. putida} typically has 5-7 flagella\cite{harwood1989flagellation}, for simplicity, our model assumes the presence of two identical flagella, as depicted in Fig. \ref{figure}.
Each flagellum is equipped with a rotary motor embedded in the cell membrane,  an elastic helical filament, and a short flexible hook that links the motor to the filament. 
Note that the hook is much more flexible than the helical filament. 

We first describe the motion of the cell body, which is neutrally buoyant and takes the shape of a capsule. 
The body surface is discretized by $n_{\text{B}}$ points and represented by two Lagrangian descriptions, ${\bf{X}}_{i}^{\text{B}}(t)$ and ${\bf{Y}}_{i}^{\text{B}}(t)$, $i = 1,...,n_{\text{B}}$. The former description ${\bf{X}}_{i}^{\text{B}}(t)$ interacts with the surrounding fluid, 
whereas the latter description ${\bf{Y}}_{i}^{\text{B}}(t)$ has no interaction with the fluid and moves as a rigid body. For each $i$, the corresponding points are linked by a stiff spring generating the following force:
\begin{equation}\label{eq:2}
{\bf F}_{i}^{\text{B}}(t) = \alpha ({\bf X}_{i}^{\text{B}} (t)- {\bf Y}_{i}^{\text{B}} (t)),
\end{equation}
where $\alpha$ is a penalty parameter that determines how tightly the two Lagranigin markers are tied together. 
This penalty force ${\bf F}_{i}^{\text{B}}(t)$ acts on ${\bf{Y}}_{i}^{\text{B}}(t)$, while $-{\bf F}_{i}^{\text{B}}(t)$ acts on the fluid. 
In addition, the spring force ${\bf F}_{i}^{\text{B}}(t)$ induces torque about the centroid of the rigid body, 
which will be described below.
The reference configuration of the rigid body is denoted by the time-independent vectors 
${\bf Z}_{i}$ satisfying $\sum_{i=1}^{n_{\text{B}}} {\bf Z}_{i}=0$. 
Then the configuration of the rigid body at time $t$,  ${\bf{Y}}_{i}^{\text{B}}(t)$, is given by
\begin{equation}\label{eq:1}
    {\bf Y}_{i}^{\text{B}}(t) = {\bf C}(t) + \mathcal{R}(t) {\bf Z}_{i}, \qquad i = 1,...,n_{\text{B}},
\end{equation}
where ${\bf C}(t)$ is the centroid of $\{ {\bf Y}_{i}^{\text{B}}(t) : i= 1,..., n_{\text{B}}\}$ and $\mathcal{R}(t)$ is a rotation matrix.

Now let ${\bf f}^{\text{B}}$ and ${\bf n}^{\text{B}}$ be the sum of all forces and torques, respectively, acting on the body
other than those generated from the coupling springs. Then the force and torque balance equations for the cell body are given, respectively, as follows:
\begin{equation}\label{eq:3}
\begin{gathered}
0 = {\bf f}^{\text{B}}(t) + \sum_{i=1}^{n_{\text{B}}} {\bf F}_{i}^{\text B}(t), 
\\
0 = {\bf n}^{\text{B}}(t) + \sum_{i=1}^{n_{\text{B}}} (\mathcal{R}(t) {\bf Z}_{i} \times {\bf F}_{i}^{\text{B}}(t)).
\end{gathered}
\end{equation}
With ${\bf X}_{i}^{\text{B}}(t)$, ${\bf f}^{\text{B}}(t)$ and ${\bf n}^{\text{B}}(t)$ known at any time, we can solve 
Eqs. (\ref{eq:1})--(\ref{eq:3}) for ${\bf C}(t)$ and $\mathcal{R}(t)$ \cite{lee2021novel}. 

Next we describe the motion of flagella based on the non-standard version of Kirchhoff rod theory \cite{lim2008dynamics}. 
Each flagellum is represented by a space curve ${\bf X}_{n}(s,t)$ and the associated orthonormal triad $\{ {\bf D}_{n}^{1}(s,t),{\bf D}_{n}^{2}(s,t), {\bf D}_{n}^{3}(s,t)\}$, $n=1,\cdots,n_{\text F}$,
where $n_{\text F}$ is the number of flagella and $s$ is a Lagrangian coordinate along the flagellum, $0 \leq s \leq L_{\text{h}}+ L_{\text{f}}$, and $L_{\text{h}}$ and $L_{\text{f}}$ are the lengths of the hook and the filament, respectively.
To construct the initial configuration ${\bf X}_n(s,0)$, we use the following reference helical flagellum:
\begin{equation}\label{eq:4}
    \hat{{\bf X}}(s)=  (r(s)\cos ( k s), r(s) \sin (k s), s),
\end{equation}
 where $k$ is the wave number, the short hook is assumed to be intrinsically straight with $r(s)=0$, $0 \leq s \leq L_{\text{h}}$, and the filament takes a spiral form with the radius $r(s) = R(1-e^{-c(s-L_{\text{h}})^2})$, $L_{\text{h}} \leq s \leq L_{\text{h}} + L_{\text{f}}$. The helical radius of the filament gradually increases to $R$. We set $c=2$ for our model organism. 
The reference flagellum  $\hat{{\bf X}}(s)$ is embedded into the cell body, positioned normal to the surface at the onset. 
At the motor point ${\bf X}_{n}(0,0)$, the unit tangent vector is aligned with ${\bf D}_{n}^{3}(0,0)$, while the principal normal and binormal vectors align with ${\bf D}_{n}^{1}(0,0)$ and ${\bf D}_{n}^{2}(0,0)$, respectively.

To describe the forces and torques of $n_{\text F}$ flagella driven rotary motors, we let ${\bf F}_{n}(s,t)$ and ${\bf N}_{n}(s,t)$, 
$n=1, \cdots, n_{\text F}$, be the internal forces and torques transmitted across a section of the flagellum, respectively, and let ${\bf f}_{n}(s,t)$ and ${\bf n}_{n}(s,t)$ be the applied force and torque densities, respectively.
The balance equations for the linear and angular momenta are given as
 \begin{equation}\label{eq:5}
     0 = {\bf f}_{n}+ \dfrac{\partial {\bf F}_{n}}{\partial s}, \qquad 0 = {\bf n}_{n}+ \dfrac{\partial {\bf N}_{n}}{\partial s} + \dfrac{\partial {\bf X}_{n}}{\partial s}\times {\bf F}_{n},
 \end{equation}
where the internal force and torque can be written in the basis of the orthonormal triad as
 \begin{equation}\label{eq:6}
 \begin{gathered}
     {\bf F}_{n} =\sum_{i=1}^{3} b_{i}\Big( {\bf D}_{n}^{i} \cdot \dfrac{\partial {\bf X}_{n}}{\partial s} - \delta_{3i} \Big){\bf D}_{n}^{i},\\
      {\bf N}_{n} =\sum_{i=1}^{3} a_{i}\Big(\dfrac{\partial {\bf D}_{n}^{j}}{\partial s} \cdot {\bf D}_{n}^{k} - \kappa_{i}\Big){\bf D}_{n}^{i},
     \end{gathered}
     \end{equation}
     where $\delta_{3i}$ is the Kronecker delta, $(i,j,k)$ is a cyclic permutation of $(1,2,3)$, and $\kappa_{i} (s)$'s describe the intrinsic curvature and twist which determine the helical radius and the pitch of the flagellum.
These constitutive relations can be derived from a variational argument of the elastic energy functional given by
\begin{equation}\label{eq:7}
    E_{n} = \dfrac{1}{2} \int \Big[
    \sum_{i=1}^{3} a_{i} \Big(\dfrac{\partial {\bf D}_{n}^{j}}{\partial s} \cdot {\bf D}_{n}^{k} - \kappa_{i} \Big)^2
    + \sum_{i=1}^{3} b_{i} \Big({\bf D}_{n}^{i} \cdot \dfrac{\partial {\bf X}_{n} }{\partial s} - \delta_{3i} \Big)^2 
    \Big] ds 
\end{equation}
where the parameters $a_1$ and $a_2$ are two bending moduli about ${\bf D}^1_n$ and ${\bf D}^2_n$, respectively, and $a_3$ is the twist modulus. The parameters $b_1$ and $b_2$ are shearing moduli and $b_3$ is the stretching modulus. 
In the limit as $b_i \rightarrow \infty$, we recover the standard  Kirchhoff rod model \cite{lim2008dynamics}.
For simplicity, we assume that $a_1=a_2$ and $b_1=b_2$ in this work. 

As a cell moves through fluid, we impose two constraints at the motor base of each flagellum 
in order to keep the flagellum attaching to and being orthogonal to the cell body at the motor points as follows:
\begin{equation}\label{eq:8}
\begin{gathered}
    \hat{\bf f}_{n}(t) = \beta ({\bf Y}_{n}^{\text F}(t) - {\bf X}_{n}(0,t)), \\
    \hat{\bf n}_{n}(t) = \gamma ({\bf D}_{n}^{3}(0,t) \times {\bf E}_{n}(t)  ), 
    \end{gathered}
    \end{equation}
where ${\bf Y}_{n}^{\text F}(0)$, $n=1,\cdots,n_\text{F}$, is the chosen surface point at which the $n$-th flagellum is attached at the onset, ${\bf E}_{n}(t)$ is the unit normal vector to the surface at the same point, and $\beta$ and $\gamma$ are sufficiently large constants to enforce the constraints. For the flagellum $n$, $\hat{\bf f}_{n}(t)$ and $\hat{\bf n}_{n}(t)$  are added to the applied force ${\bf f}_{n}(0,t)ds$ and torque ${\bf n}_{n}(0,t)ds$ in equation (\ref{eq:5}), respectively, while $-\sum_{n=1}^{n_\text{F}}\hat{\bf f}_{n}(t)$ and $-\sum_{n=1}^{n_\text{F}}\hat{\bf n}_{n}(t)$ are added to the total external force ${\bf f}^{\text B}$ and torque ${\bf n}^{\text B}$ acting on the cell body in equation (\ref{eq:3}), respectively.

In order to drive the flagellar rotation by a rotary motor, we discretize each flagellum denoted by ${\bf X}_{n}(j \Delta s,t)$, $j=0,...,M-1$, 
where $\Delta s$ is the meshwidth along the flagellum and $M$ is the number of material points of each flagellum. 
Then we apply to the ghost point ${\bf X}_{n}(-\Delta s/2,t)$ the following constant torque in the normal direction ${\bf E}_n(t)$: 
\begin{equation}\label{eq:9}
    {\bf N}_{n}(-\Delta s/2,t) = -\tau_{n}{\bf E}_{n}(t),
\end{equation}
where $\tau_n$ determines the direction of rotation and the magnitude of motor torque. 
The motor rotates counterclockwise (CCW) when $\tau_{n}<0$ and clockwise (CW) when $\tau_{n}>0$.  
Note that the total torque $\sum_{n=1}^{n_{\text F}}{\bf N}_{n}(-\Delta s/2,t)$ is applied to the cell body, which results in 
counterrotation of the cell body.

Lastly,  the cell dynamics is coupled to the surrounding fluid governed by the viscous incompressible Stokes equations \cite{olson2013modeling}:
\begin{equation}\label{eq:10}
    0 = -\nabla p + \mu \Delta {\bf u} + {\bf g}, \qquad 0 = \nabla \cdot {\bf u},
\end{equation}
where the fluid velocity ${\bf u}({\bf x},t)$ and the fluid pressure $p({\bf x},t)$ are unknown variables as functions of the Carterian coordinates ${\bf x}$ and time $t$, and $\mu$ is the fluid viscosity. The external force density ${\bf g}$ applied to the fluid by the immersed cell is given as
\begin{align}
    \begin{aligned}
        {\bf g}({\bf x}, t) = &\sum_{n=1}^{n_{\text F}} \Big[ \int_{0}^L \big(-{\bf f}_{n}(s,t) \big) \psi_{\varepsilon}({\bf x} - {\bf X}_{n}(s,t))ds \\
        & \quad  + \dfrac{1}{2} \nabla \times \int_{0}^L  \big(-{\bf n}_{n}(s,t) \big) \psi_{\varepsilon}({\bf x} -{\bf X}_{n}(s,t) ) ds \\
        & \quad + \int_{0}^L (- {\bf f}_{n}^{\text r}(s,t)) \psi_{\varepsilon}({\bf x} - {\bf X}_{n}(s,t))ds \\
        &  \quad  + \sum_{i=1}^{n_{\text{B}}}\int_{L'}^L (- {\bf f}_{n,i}^{\text{r}}(s,t)) \psi_{\varepsilon}({\bf x} - {\bf X}_{n}(s,t))ds 
        \\
        & \quad  + \sum_{i=1}^{n_{\text{B}}} \int_{L'}^L {\bf f}_{n,i}^{\text r} (s,t) \psi_{\varepsilon}(
        {\bf x} - {\bf X}_{i}^{\text B}(t))ds \Big] \\
        & + \sum_{i=1}^{n_{\text{B}}} (-{\bf F}_{i}^{\text B} (t))  \psi_{\varepsilon}(
        {\bf x} - {\bf X}_{i}^{\text B}(t)),
    \end{aligned}
\end{align}
where $L=L_{\text f} + L_{\text h}$.  We set $L'=5\Delta s$ so that the repulsive force does not activate near the motor. The first two terms correspond to the forces and torques generated from the flagella and the last term is the force acting on the fluid by the cell body. The repulsive force functions ${\bf f}_{n}^{\text r}(s,t)$ and ${\bf f}_{n,i}^{\text r}(s,t)$ prevent contacts between flagella themselves, and between the flagella and the cell body, respectively, given as
\begin{widetext}
\begin{align}
    \begin{aligned}
{\bf f}_{n}^{\text r} (s,t) =& \int C \Big[ \max \Big( 1 - \dfrac{\| {\bf X}_{n}(s,t)- {\bf X}_{n}(s',t)\|}{D} ,0 \Big)\Big]  \dfrac{{\bf X}_{n}(s,t) - {\bf X}_{n}(s',t)}{\| {\bf X}_{n}(s,t) - {\bf X}_{n}(s',t)\| }ds'
\\
&+\sum_{\substack{n'=1 \\ n'\ne n}}^{n_{\text F}}\int C \Big[
   \max \Big( 1 - \dfrac{\| {\bf X}_{n}(s,t)- {\bf X}_{n'}(s',t)\|}{D},0 \Big)\Big] 
   \dfrac{{\bf X}_{n}(s,t) - {\bf X}_{n'}(s',t)}{\| {\bf X}_{n}(s,t) - {\bf X}_{n'}(s',t)\| }ds',
   \\
   {\bf f}_{n,i}^{\text r}(s,t)&=C
\Big[ 
   \max \Big( 1 - \dfrac{\| {\bf X}_{n}(s,t)- {\bf X}_{i}^{\text B}(t)\|}{D},0 \Big)\Big]
   \dfrac{{\bf X}_{n}(s,t) - {\bf X}_{i}^{\text B}(t)}{\| {\bf X}_{n}(s,t) - {\bf X}_{i}^{\text B}(t)\| }.
   \end{aligned}
\end{align}
\end{widetext}
The constant $C$ determines the strength to keep the distance between flagella and the cell body with the minimum distance $D.$ Here, a regularized Dirac delta function $\psi_\epsilon({\bf r})$ is defined by
   \begin{equation}
       \psi_{\varepsilon} ({\bf r}) = \dfrac{15 \varepsilon^{4}}{8 \pi (\| {\bf r}\|^{2}  + \varepsilon^{2})^{7/2}}
   \end{equation}
satisfying $\int_{\mathbb{R}^{3}} \psi_{\varepsilon} ({\bf r}) d{\bf r}=1$. The size of $\varepsilon$ determines the effective radius of the immersed boundaries \cite{lee2018bacterial}.

In the end, the motion of the flagella and the cell body are described by
\begin{align}
    \begin{aligned}
        \dfrac{\partial {\bf X}_{n}}{\partial t}(s,t) &= {\bf u}\big( {\bf X}_{n}(s,t),t \big), \quad n=1, \cdots, n_{\text F},\\
        \qquad \dfrac{\partial {\bf X}_{i}^{\text B}}{\partial t}(t) &= {\bf u}\big({\bf X}_{i}^{\text B}(t),t \big), \quad i=1, \cdots, n_{\text B},\\
        \dfrac{\partial {\bf D}_{n}^{j}(s,t)}{\partial t} &= {\bf w}\big( {\bf X}_{n}(s,t),t \big) \times {\bf D}_{n}^{j}(s,t), \\
       & \qquad \qquad j=1, 2, 3, \;\ \text{and } \;\ n=1, \cdots, n_{\text F}
    \end{aligned}
\end{align}
where the angular velocity of the fluid ${\bf w}$ is defined as
\begin{equation}
    {\bf w}({\bf x},t) = \dfrac{1}{2} \nabla \times {\bf u}({\bf x},t).
\end{equation}
The related parameters are listed in Supplementary 
Table S1, and more details about the method can be found in the literature.\cite{lee2021novel,lim2008dynamics, lim2012fluid,  olson2013modeling}

\section{\label{sec:results}Results}

We classify the swimming modes of lophotrichously flagellated cells with two flagella based on various combinations of the two motor torques, specifically considering the direction of motor rotation and the magnitude of the torque. 
Among the swimming modes, there are cases in which two flagella form a stable bundle, a crucial aspect for many biological processes in flagellated bacteria. We investigate the distinguishing characteristics of stable bundles from one another.
Lastly, we examine the effect of material properties of the filament and the hook, particularly when two motors turn in the same direction with equal magnitudes of torques.

\subsection{\label{subsec:result1} Classification of swimming modes determined by motor torques}

\begin{figure*}[t!]
	\centering
\includegraphics[width=0.9\textwidth]{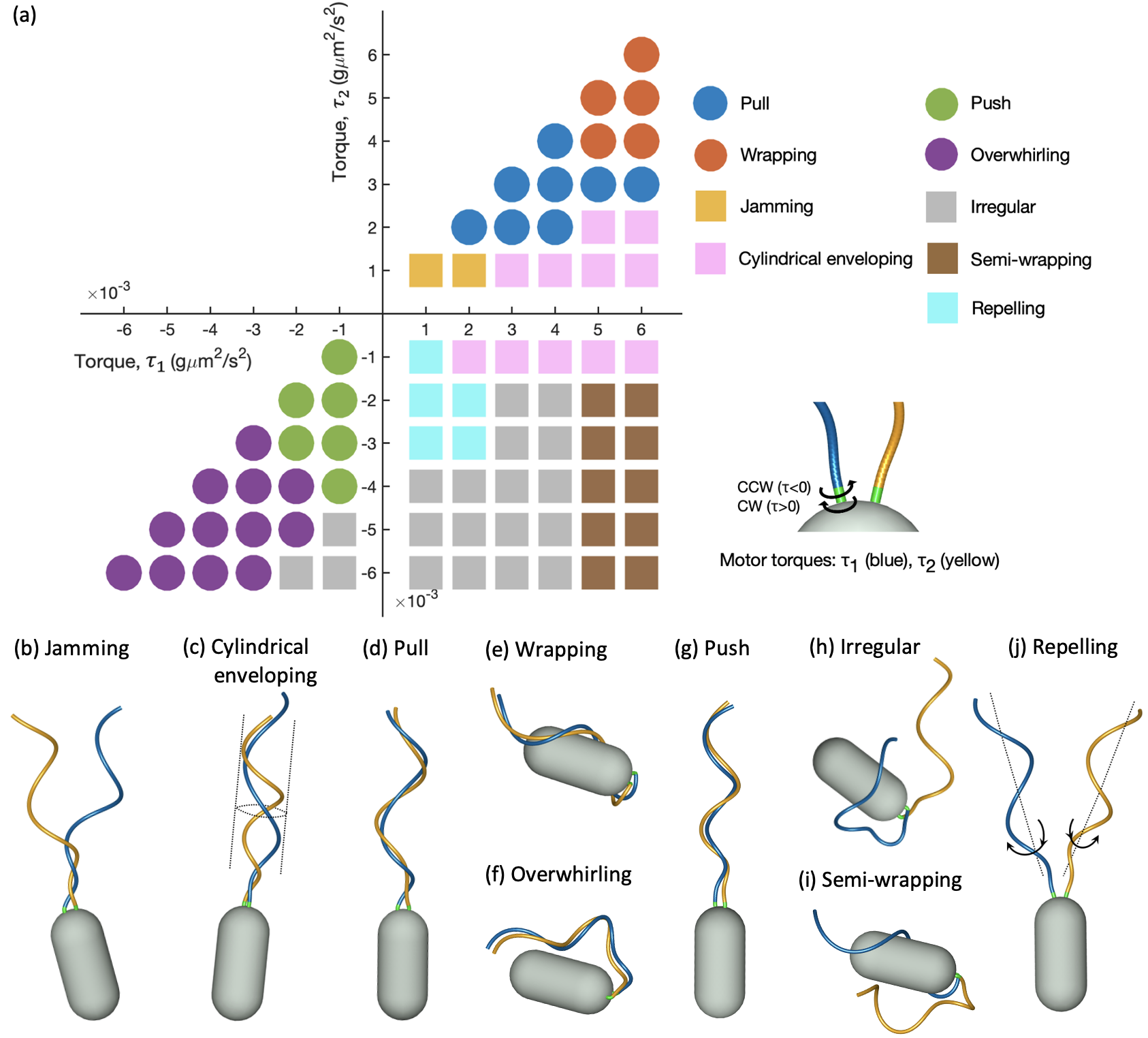}
	\caption{Modes of swimming motility determined by motor torques. 
 (a) Classification of swimming modes depending on applied motor torques $\tau_{1}$ and $\tau_{2}$. 
Each motor rotates either CW (positive values of torque) or CCW (negative values of torque) and the hydrodynamic interaction of a cell results in various modes of motility.
Nine different modes are displayed by different colors of markers: 
(b) jamming, (c) cylindrical enveloping, (d) pull, (e) wrapping, (f) orverwhirling, (g) push, (h) irregular, (i) semi-wrapping, and (j) repelling modes. 
The flagella take the form of a bundle in modes (d)-(g) but not for the rest of the modes. 
Moreover, the bundle in pull and push modes are the only stable twirled bundles, while the bundle in wrapping and overwhirling modes
buckles and becomes unstable. Circles and squares correspond to bundled and unbundled forms, respectively.
(Multimedia available online).
}
 \label{fig:diff_torque}
\end{figure*}

As flagellar motors rotate and generate torques which drive the flagellar rotation, accompanied by counterrotation of the cell body,
hydrodynamic interactions between the flagellar filaments are crucial for the cell's motility.
Many factors such as geometry and material properties of the cell also contribute to the cell's swimming patterns.
In particular, the magnitude of the applied torque by individual motors and the direction of motor rotation are important factors that determine the various modes of motility. 
Figure \ref{fig:diff_torque} shows a classification of swimming modes, in which the two flagellar motors rotate with various combinations of 
torque magnitude and rotational direction, and hence the cell attains a steady state motion.
We consider two torque parameters, $\tau_1$ and $\tau_2$,
with each one ranging from $-0.006$ g$\mu$m$^2$/s$^2$ to $0.006$ g$\mu$m$^2$/s$^2$. 
Here, the positive and negative values of those parameters correspond to CW and CCW, respectively, 
see the inset of Fig. \ref{fig:diff_torque}(a). The rest of the parameter values are being held fixed as in Supplementary 
Table S1.
Note that simulation results in Fig. \ref{fig:diff_torque} are symmetric about the line $\tau_2=\tau_1$. 

When  both motors turn CW, i.e., $\tau_1>0$ and $\tau_2>0$, corresponding to the first quadrant in Fig. \ref{fig:diff_torque}(a), 
there are  four dynamical regimes of the motion depending on torque magnitudes: jamming, cylindrical enveloping, pull, and wrapping. 
See Fig. \ref{fig:diff_torque}(b-e) (Multimedia available online).
{\it Jamming} is the motion in which two flagella start coiling around each other but being stuck together before completing a flagellar bundle. 
This motion occurs when both torques are very small so that the given torques are not strong enough to drive flagellar rotation.
{\it Cylindrical  enveloping} occurs when one torque is very small while the other torque is relatively large, wherein two flagella keep rotating within the cylindrical surface but fail to form a bundle. As both torques increase to some extent, two flagella form a bundle
which is aligned with and rotate about the body axis, and steadily move the cell backward,
called {\it pull}. When both torques further increase, two flagella first form a bundle, then buckle and coil around the cell body while rotating. This phenomenon is called {\it wrapping}.
The pull and wrapping modes are frequently observed in the motion of {\it P. putida}, 
and there exists a critical threshold of torque magnitude that separates pull from wrapping.

When both motors turn CCW, $\tau_1<0$ and $\tau_2<0$, corresponding to the third quadrant in Fig. \ref{fig:diff_torque}(a), 
there are three swimming modes: push, overwhirling, and irregular modes. 
See Fig. \ref{fig:diff_torque}(f-h) (Multimedia available online).
When both $|\tau_1|$ and $|\tau_2|$ are relatively small, both flagella pump the fluid behind the cell,
attract each other and eventually synchronize in phase forming a bundle. This motion exerts thrust on the
cell body and thus moves the cell forward, called {\it push}. 
The push mode is a typical forward run mode observed in many bacteria such as {\it E. coli}, {\it Vibrio A}, and {\it P. putida}.
If both torque magnitudes become larger, the quickly formed bundle buckles drastically and rotate 
about the axis of the cell body, which is called {\it overwhirling}. 
Note that there exists a threshold of torque that separates overwhirling from push mode.
The {\it irregular} mode occurs when one torque magnitude is small, but the other one is large.
The large magnitude of the torque leads the flagella to overwhirl and to work independently. 
This mode of motility becomes more prevalent when the motors spin in opposite directions.

Lastly consider the cases where two motors turn in the opposite directions with $\tau_1>0$ and $\tau_2 <0$.
The flagellar interaction results in four different modes: cylindrical evenloping, semi-wrapping, repelling, and irregular modes.
See Fig. \ref{fig:diff_torque}(c, h-j) (Multimedia available online).
Cylindrical enveloping occurs in general when $|\tau_2|$ is very small and the other one is large as previously observed.
{\it Repelling} is the motion in which  one flagellum spins CCW to propel fluid behind the cell, 
while the other flagellum rotates CW to move fluid toward the cell, resulting in the two flagella remaining separated while in rotation. 
This phenomenon occurs when the magnitudes of both torques are  
at similar levels, yet they are not excessively high. 
As the magnitude of CCW-spinning motor torque, $|\tau_2|$, increases further while the other remains within a certain range,
the two flagella initially repel each other due to the opposite directional rotation but then exhibit an erratic movement. 
This irregularity is dominant in the region where two motors turn in opposite directions.
Finally, {\it semi-wrapping} refers to the motion in which one flagellum wraps around the cell body 
while the other flagellum overwhirls, particularly when $\tau_1$ is maintained at a larger value.  

Supplementary Figure S1  
depicts the total elastic energy along the flagella over time for seven representative swimming modes which is calculated by Eq. (\ref{eq:7}). Time evolution 
of energies is illustrated after the stable motion is reached. Here, bending and twist energies are dominant, whereas the shearing and stretching energies are negligible. Notably, the wrapping and overwhirling modes experience buckling instability, leading
to substantial deformations and higher energy consumption in bending and twist. 
Overall, there are four types of bundled forms: pull and wrapping when both motors turn CW, and push and overwhirling when both motors turn CCW. 
The torque level critically determines whether the bundle remains standing or becomes buckled. 
 When the magnitude of both motors is relatively small, the bundle stays positioned behind the cell body, aligning with its axis. However, 
 as the magnitude increases, the bundle buckles due to the flexible hook. In the following section,
 we will delve deeper into the dynamics of two stable bundles: pull and push modes.

\subsection{\label{subsec:stability}Stable bundles: push and pull modes}

When multiple motors rotate synchronously, the flagella form a stable bundle,
aligning with the collective behavior resembling that of a single flagellum \cite{hintsche2017polar}. 
Our simulations corroborate this observation, 
prompting a detailed exploration of flagellar dynamics within stable bundles,
particularly in the context of pull and push modes.

\begin{figure*}[t!]
	\centering
\includegraphics[width=0.9\textwidth]{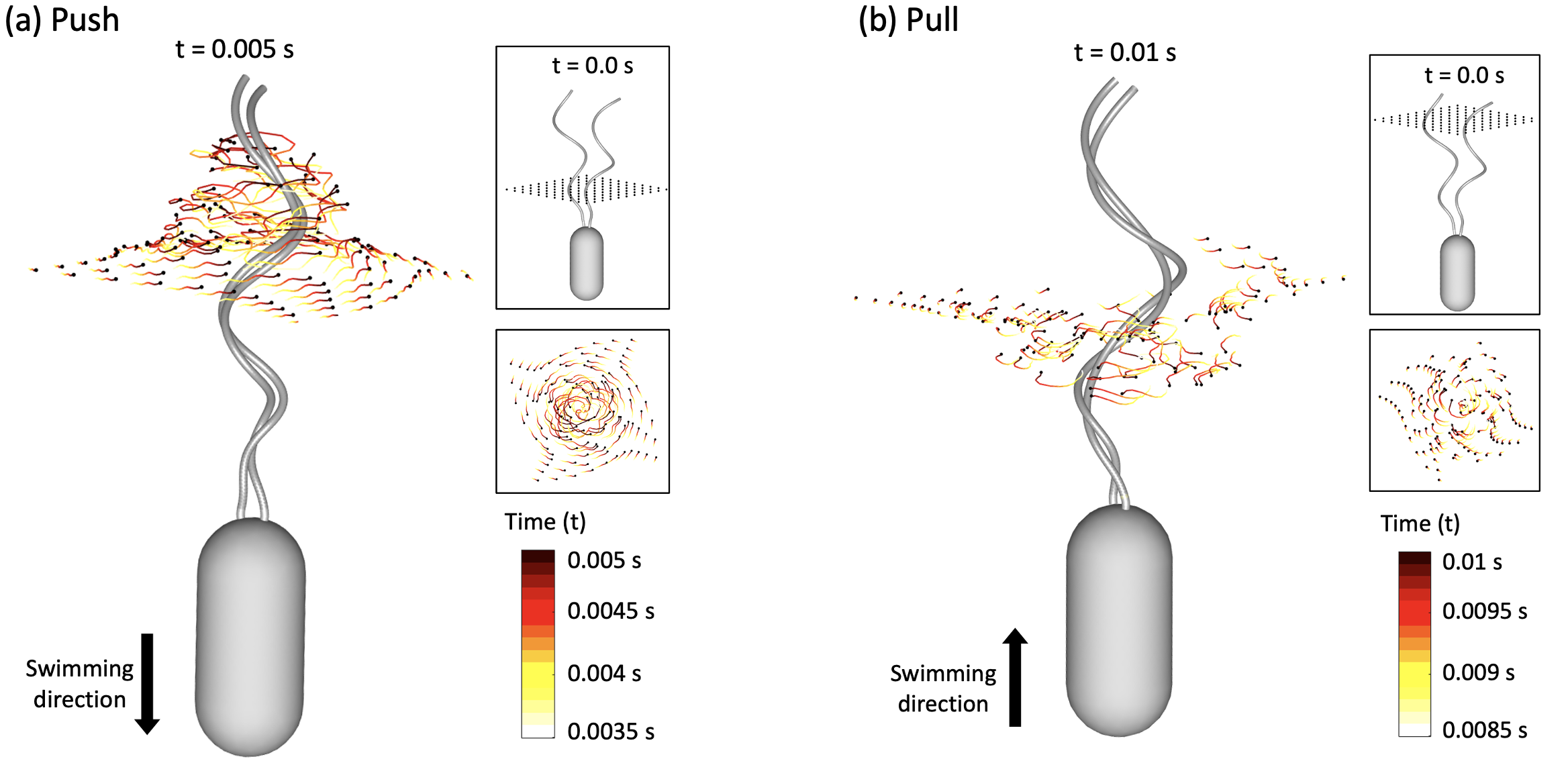}
	\caption{
 Snapshots of the swimming cells with the fluid markers when the synchronized motors rotate in CCW (a) and CW (b).
 Initially, the markers are uniformly distributed, as shown in each top inset. 
 Over time, the markers' trajectories trace the flow around the flagellar bundle, 
 and their projection on the plane perpendicular to the helical axis is displayed in each second top inset.
 The color bar illustrates the time evolution of markers' trajectories for a duration of 0.0015 s. 
 The magnitude of all the applied motor torque is set to be 0.00275  $\text{g}\mu\text{m}^2/{\text s}^2$.
(Multimedia available online).}
 \label{fig:marker}
\end{figure*}

Figure \ref{fig:marker} shows snapshots of the swimming cells with the passive fluid markers 
flowing around the flagellar bundle when the two synchronized motors rotate either CCW (left panel) or CW (right panel).
At onset, the fluid markers are uniformly distributed on the rectangular plane, as shown in each top inset, 
and then they trace flow dynamics around the flagellar bundle. 
The colored paths traced by fluid markers depict temporal trajectories,
with the leading position marked in black at the present time.
Marker trajectories at given times are projected onto the plane perpendicular
to the helical axis, as shown in each second middle inset.
Similar to the behavior of cells with a single flagellum, a stable bundle rotating CCW 
pumps the fluid behind the cell body, thus causing the cell to move forward, whereas a stable bundle rotating CW pumps the fluid toward the cell body, resulting in a backward run (Multimedia available online).
In both cases, the bundle is left-handed; however, the helical pitch and radius become larger in pull mode than in push mode.

\begin{figure*}[t!]
	\centering
\includegraphics[width=0.9\textwidth]{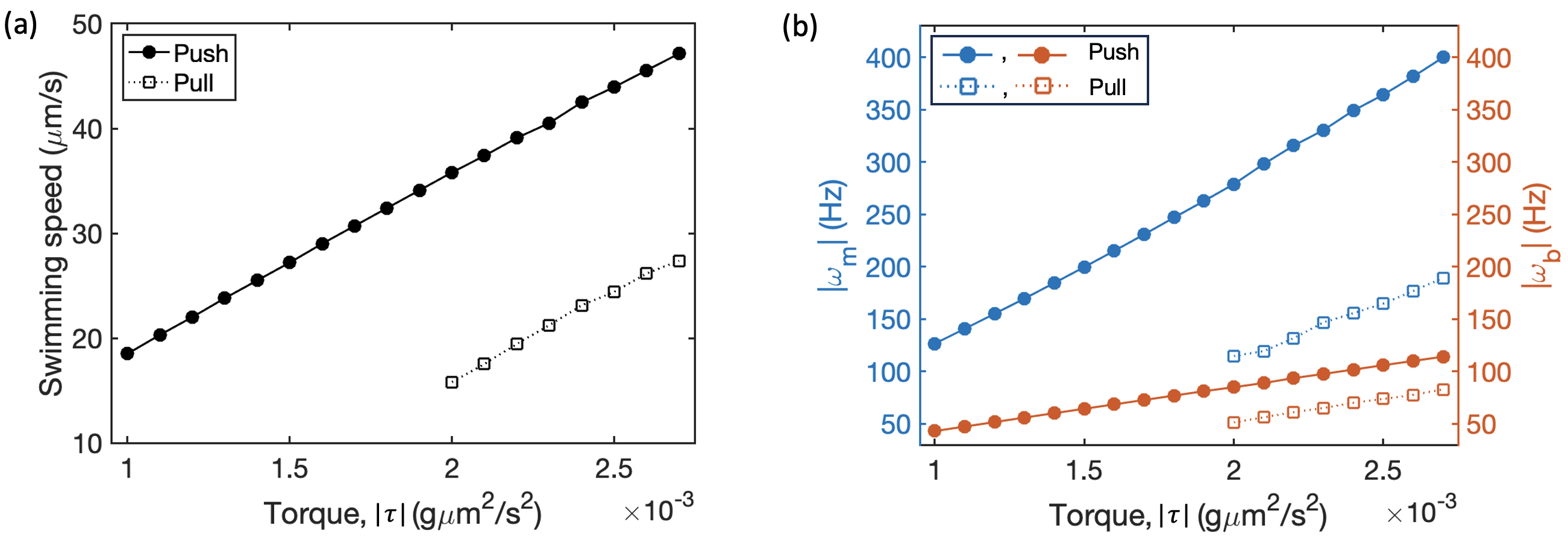}
	\caption{ 
	Linear relationship between the applied torque and 
 swimming speed (a), the average of individual motor rotation rate $|\omega_{\text{m}}|$
 and the body rotation rate $|\omega_{\text{b}}|$ (b) during the stable push and pull modes.
}
 \label{fig:bundle speed rotation rate}
\end{figure*}

Figure \ref{fig:bundle speed rotation rate} shows that 
the swimming speed and rotation rates of the motors and the cell body increase linearly 
with the applied motor torque in both pull and push modes. This trend agrees with
the numerical observations reported in Park et al.\cite{park2022modeling},
in which a single flagellum was the subject of investigation. 
The push mode generally yields higher magnitudes of rotation rates for both the flagella and the cell body when subjected to the same torque magnitude. This phenomenon contributes to faster swimming speeds compared to the pull mode.
Notably, this linear relationship holds regardless of the number of flagella involved.
It is pertinent to note that the swimming speeds of the push and pull modes of 
the wild type {\it P. putida} are approximately 25$\mu$m and 30$\mu$m, respectively, when the flagellar rotation rate reaches around 200 Hz~\cite{hintsche2017polar}, which agrees with our simulations.

\begin{figure*}
	\centering
\includegraphics[width=0.9\textwidth]{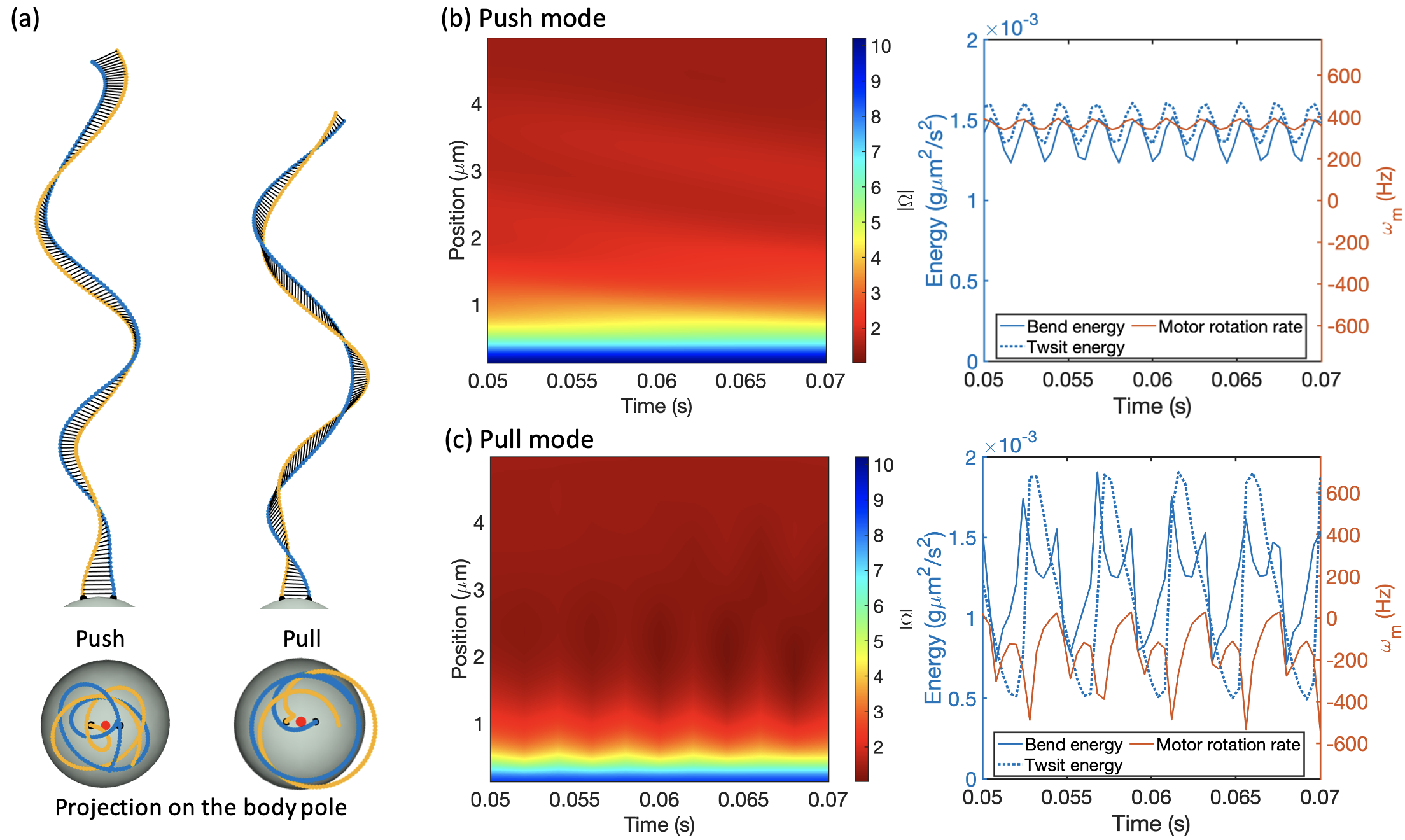}
	\caption{
Comparison of bundle formation between push and pull modes. The applied motor torque for each mode remains fixed at -0.0025 $\text{g}\mu\text{m}^2/{\text s}^2$
for the push mode and 0.0025 $\text{g}\mu\text{m}^2/{\text s}^2$ for the pull mode. (a)
Comparison of intertwining levels between the push (left) and pull (right) modes. Two flagella of each mode
are depicted by helical curves in yellow and blue, while line segments illustrate connections between
corresponding material points of the two flagella for visualization purposes. The bottom insets exhibit the projection of the respective flagella onto the cell body, with the pole marked by a red dot.
(b,c) Time evolution of torsion  (left), bending and twist energies, and motor rotation rate (right) of one flagellum for push (b) and pull (c). The color bars show the magnitude of torsion values $\Omega$ along the flagellum over time.
The right panels show the bending and twist energies and the motor rotation rate.
The average values of bending and twist energies are 0.0014 $\text{g}\mu\text{m}^2/{\text s}^2$and 0.0015 $\text{g}\mu\text{m}^2/{\text s}^2$ for the push mode, respectively, and 0.0012 $\text{g}\mu\text{m}^2/{\text s}^2$ and 0.0011 $\text{g}\mu\text{m}^2/{\text s}^2$ for the pull mode, respectively.
}
 \label{fig:bundle}
\end{figure*}

The difference between the push and pull modes becomes evident in their configurations. 
Figure \ref{fig:bundle}(a) presents the degree of intertwining between two flagella for the two modes
by connecting the corresponding material points of the flagella. It is apparent that in the pull mode,
the flagella tend to be more intertwined, while the push mode demonstrates a phase-locked state. 
Furthermore, in the pull mode, both helical pitch and radius become greater than the given intrinsic values,
whereas they become slightly lower in the push mode.

To further analyze the deformation of one flagellum in each mode, we compare evolutions of the torsion and energy between push and pull modes, see Fig. \ref{fig:bundle}(b,c). The torsion value is evaluated as $\Omega=(\partial {\bf D}^{1}/\partial s) \cdot {\bf D}^{2}$, where ${\bf D}^1$ and ${\bf D}^{2}$  represent orthonormal vectors along the flagellum. Bending and twist energies are calculated using Eq. (\ref{eq:7}). 
In the push mode, the flagellum\rq{}s torsion exhibits slightly higher magnitudes than in the pull mode as the motor rotates. This indicates that each flagellum in push mode tends to twist more.
The energies and motor rotation rate in the push mode evolve sinusoidally over time within a narrow range.
However, in the pull mode, they fluctuate across a wide range of values, displaying dynamic changes. 
Since CCW and CW rotations of flagella induce fluid pumping in opposite directions, 
which also leads to wave propagation in corresponding opposite directions,
these contrasting fluid flows cause distinct deformation in the flagellar configuration and affect the performance of the motors.
In particular, our simulation demonstrates that the motor\rq{}s performance differs significantly, even when the same magnitude of torque is applied.
Note that shear and stretch energies remain negligible compared to the twist and bending energies throughout, 
regardless of direction of motor rotation.

\subsection{Effect of flagellar properties on swimming modes}

The material property of flagella, particularly their flexibility plays a significant role in bundle formation.  
We investigate the influence of flagellar flexibility on flagellar bundling by examining the steady states
of swimming motion. Specifically, we explore the effect of factors such as the bending modulus of the filament,
the presence of the hook, and the applied torque on these steady states. 

\begin{figure*}
	\centering
\includegraphics[width=0.9\textwidth]{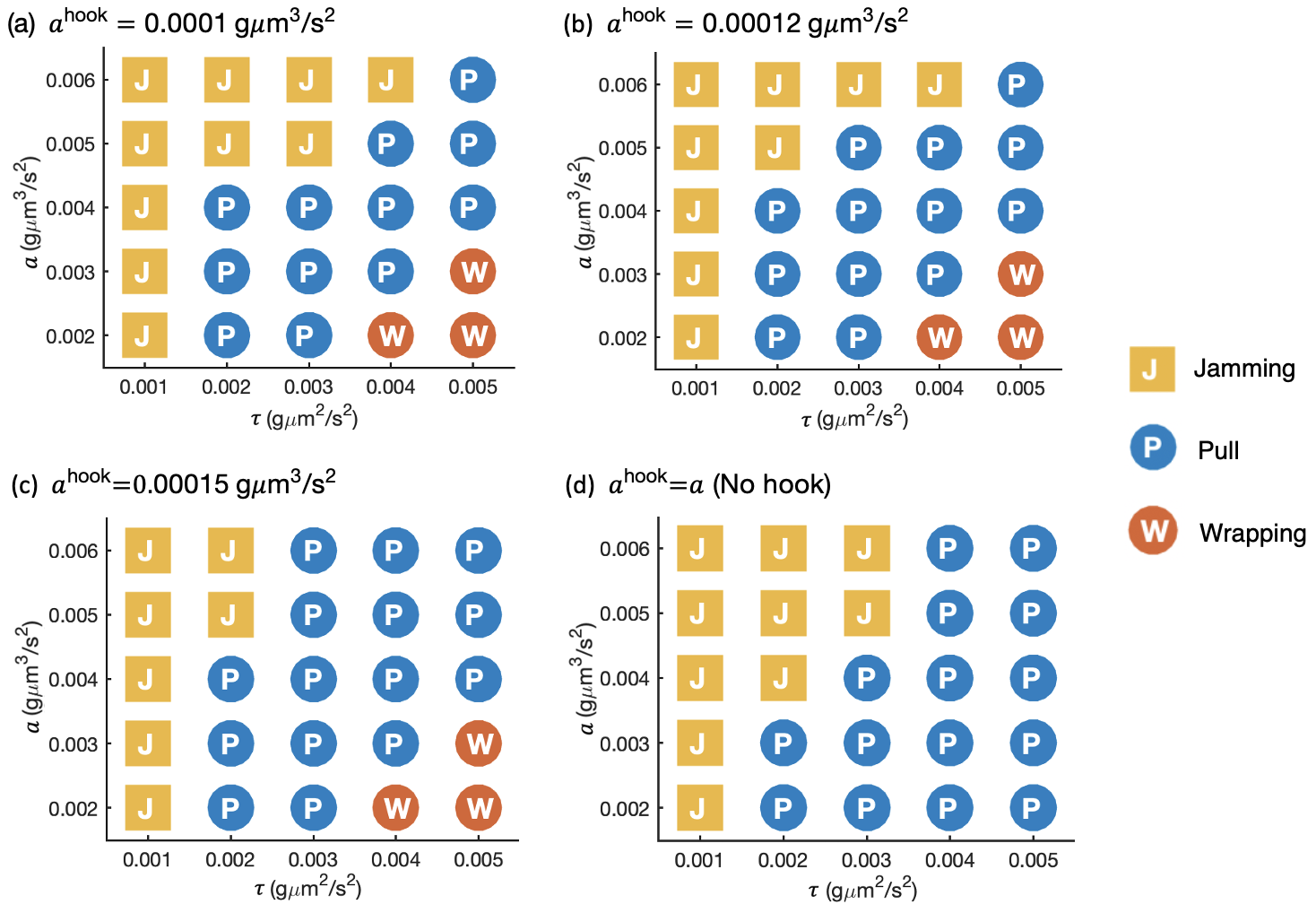}
	\caption{ 
Steady states of swimming motion when both motors rotate clockwise. Three different modes are attained, depending on the bending modulus of the filament ($a$) and the applied motor torque ($\tau$) in the presence (a-c) and absence (d) of the hook. Both motors take the same value of $\tau$.
Squares and circles distinguish failure and success in forming stable bundles, respectively.
 }\label{fig:hook effect_pull}
\end{figure*}

When both motors rotate CW with equal torque, the formation of a bundle requires a minimum torque threshold, as shown in Fig. \ref{fig:hook effect_pull}. This threshold increases with the stiffening of the flagellar filament. 
In the presence of the hook, the stiffer hook reduces the minimum torque threshold required for bundle formation 
as the filaments become more rigid, owing to enhanced transmission of motor torque along the filament. 
In the bundling cases, there is also a critical torque size that separates wrapping from pull for each bending modulus of the filament, which remains consistent irrespective of the hook's compliance, see Fig. \ref{fig:hook effect_pull}(a-c).
However, without the hook, the wrapping mode is not observed within the current range of the bending modulus of the flagellar filament and the applied motor torque. This observation underscores the pivotal role of the compliant hook in deforming a bundle for wrapping motion, consistent with the findings in Park et al. \cite{park2022modeling} with a single flagellum.

\begin{figure*}
	\centering
\includegraphics[width=0.9\textwidth]{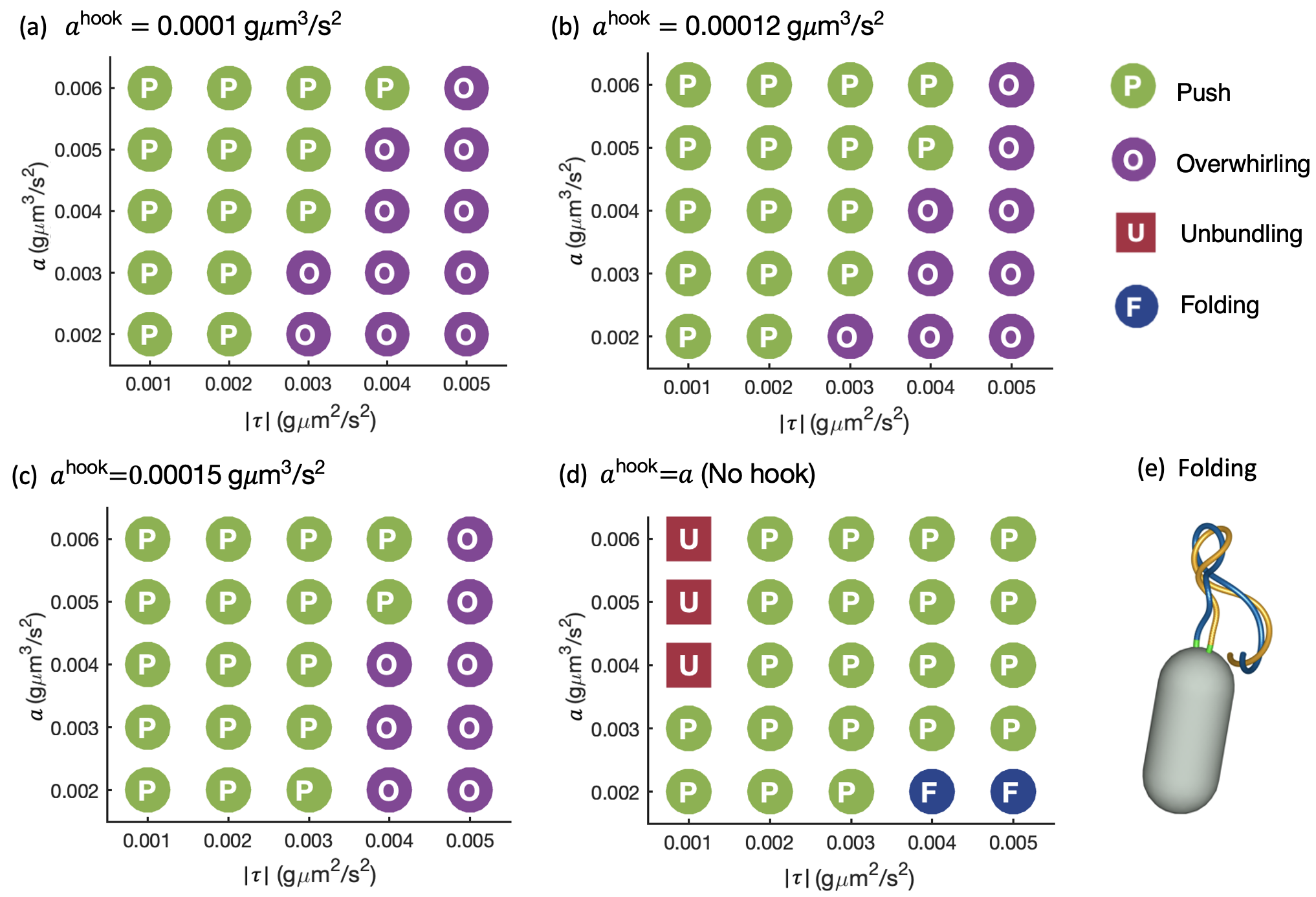}
	\caption{ 
 Steady states of swimming motions when both motors rotate counterclockwise and take the same value of $\tau$. 
 Two different modes are observed as the bending modulus of the filament ($a$) and the applied motor torque ($\tau$) vary in the presence (a-c) of the hook.  In the absence of the hook (d), two new motions emerge: unbundling and folding (e).
 }\label{fig:hook effect_push}
\end{figure*}

When both motors rotate CCW with equal torque, the flagellar bundle exhibits a push mode for low applied torque and an overwhirling mode for high torque. The torque threshold for this transition is elevated as the bending modulus of the filament or the hook increases, as illustrated in Fig. \ref{fig:hook effect_push}(a-c). Note that the hook still maintains greater flexibility compared to the filament. However, in the absence of the hook in Fig. \ref{fig:hook effect_push}(d), two new motions emerge: unbundling and folding. In the former case, two flagella remain separate, maintaining a certain distance between them while rotating independently, thereby failing to form a bundle. This motion occurs when the torque is very small, yet the filament is sufficiently rigid, resulting in weaker hydrodynamic interaction. In the latter case, two flagella rapidly form a bundle; however, due to excessive torque and the absence of the hook, the formed bundle buckles midway rather than near the motor, see Fig. \ref{fig:hook effect_push}(e).

\section{Summary and discussion}

We consider a mathematical model of the lophotrichous bacterium {\it P. putida} consisting of a rigid rod-shaped cell body with two identical flagella
positioned near one end of the cell body.  For simplicity, these two flagella are embedded normal to the cell surface and are arranged symmetrically about the body axis, and
more flagella can be attached to this numerical model organism. We primarily classify the swimming modes under different conditions of applied torque, direction of motor rotation, and flagellar elasticity. 
Moreover, we scrutinize stable flagellar bundles, particularly focusing on push and pull modes.

Despite the close proximity of flagellar motors and the potential for bundle formation in both
clockwise and counterclockwise motor rotation directions, our model reveals previously unobserved
swimming modes in multiflagellated bacteria.
These modes can be categorized into two groups based on whether the flagella are formed as a bundle or not. 
Representative stable bundle modes include pull, push, wrapping and overwhirling, which are distinguished by the torque threshold and rotational direction. 
Our simulations demonstrate that stable bundles can generally form when all motors rotate in the same direction and their torque magnitudes are within
a similar range, except in the case of the jamming mode; otherwise, they fail to form a bundle. 
It was reported that the majority of flagellar motors in {\it P. putida} undergo synchronous reversal
of motor rotation direction during a transition between swimming modes \cite{hintsche2017polar}.
Disruption of motor synchronization can lead to the failure of stable bundle formation, which may result in alterations in swimming direction and may also interrupt chemotactic
behavior of bacteria.

\begin{figure*}
	\centering
\includegraphics[width=0.9\textwidth]{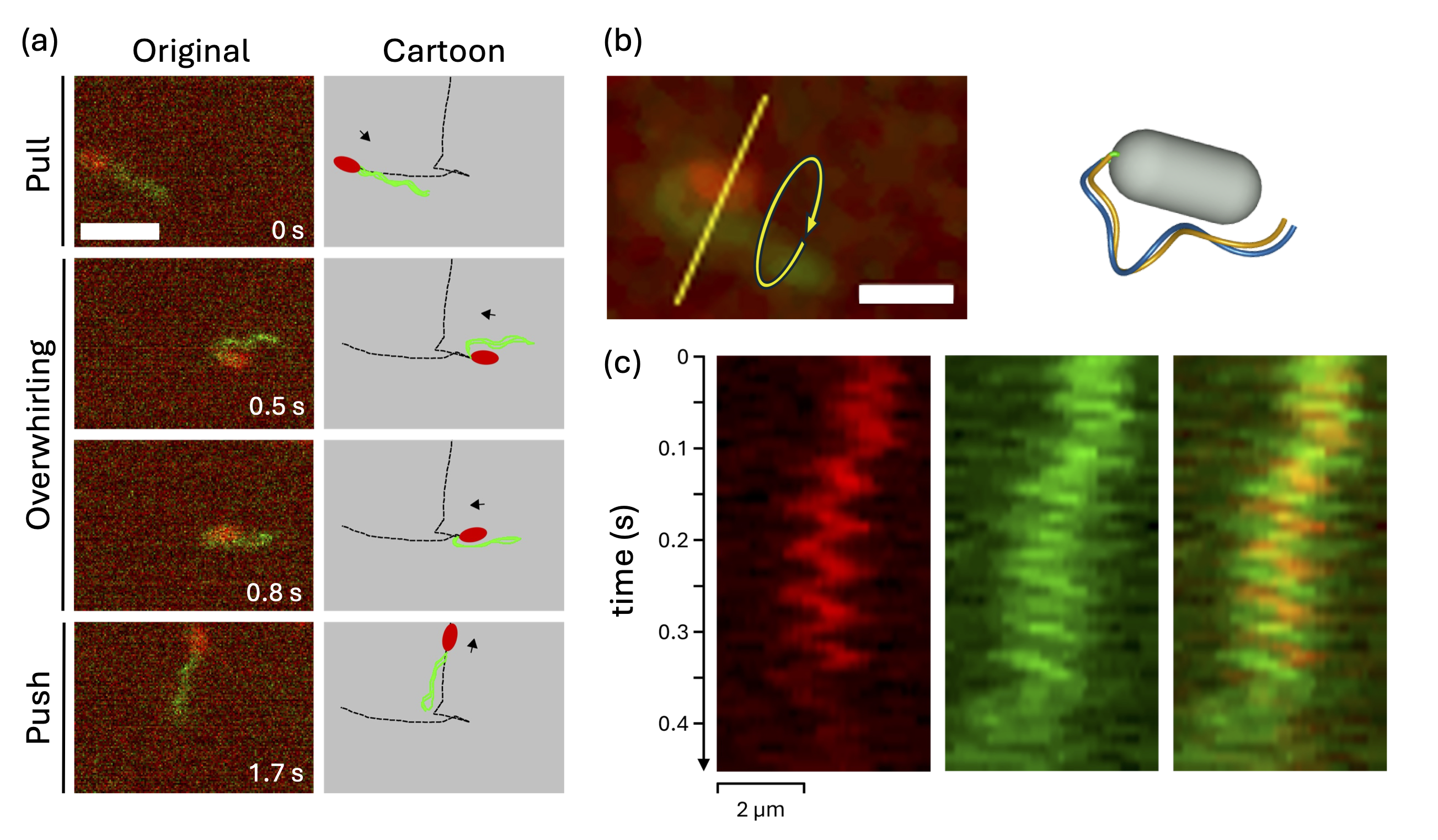}
	\caption{
 Experimental evidence of overwhirling mode.
 (a)~Time-lapse sequence of dual-color fluorescence images (left) and corresponding cartoon representations (right) of a {\it P.~putida} cell undergoing transitions from pull to overwhirling and from overwhirling to push mode (Multimedia available online).
 The cell body is displayed in red and the flagellar bundle in green; scale bar 5$\mu$m.
 (b)~Smoothed fluorescence image of a {\it P.~putida} cell in the overwhirling mode (left; scale bar 2$\mu$m) (Multimedia available online) and corresponding snapshot taken from a numerical simulation (right) (Multimedia available online).
 (c)~Space-time plot taken along the yellow line in panel~(b) showing, from left to right, the red and green fluorescence channels and an overlay.
}\label{fig:exp_pull_overwhirl_push}
\end{figure*}

Among the various swimming modes, our findings indicate that three primary swimming modes in bundled forms (pull, push, and wrapping) are most efficient in terms of swimming speed.
The overwhirling mode, however, shows relatively small displacement as compared to the other bundled forms. 
This discrepancy may explain why the overwhirling mode is not commonly observed in natural systems.
Here, we report a rare observation of a {\it P.~putida} cell in overwhirling mode, see Fig.~\ref{fig:exp_pull_overwhirl_push}.
The cell undergoes a transition from pull to overwhirling mode and, after approximately 0.5~s, switches from overwhirling to push mode, see panel~(a) (Multimedia available online).
During the episode in overwhirling, the cell exhibits only little displacement, while the entire flagellar bundle is rotating around the cell body, see arrow in the left panel of (b) (Multimedia available online).
As a result, the cell body and flagellar bundle orbit around their common center of mass, which appears as a periodic wiggling when seen in the plane of imaging.
A space-time plot recorded along the cross-sectional line in panel~(b) illustrates the wiggling motion, where the cell body and bundle proceed in antiphase, as shown by the overlay of both fluorescence channels in panel~(c). A similar sequence of pull, overwhirling, and push modes can be also observed in numerical simulations, see the right panel in (b) (Multimedia available online). Upon a simultaneous reversal of both motors from CW to CCW rotation, a cell that initially swims in pull mode will directly switch to the overwhirling mode, if the CCW torque of both motors is sufficiently large. A subsequent decrease in the CCW torque will then initiate a transition from overwhirling to push mode. Note also that in the simulation, a turning angle of around 90$^\circ$ is observed during the transition from overwhirling to push mode. In contrast to the common directional reversals in the trajectories of swimming {\it P.~putida} cells, turning angles of 90$^\circ$ provide an additional degree of freedom that may be beneficial to efficiently spread and explore the environment. 

\begin{figure*}
	\centering
\includegraphics[width=1.5\columnwidth]{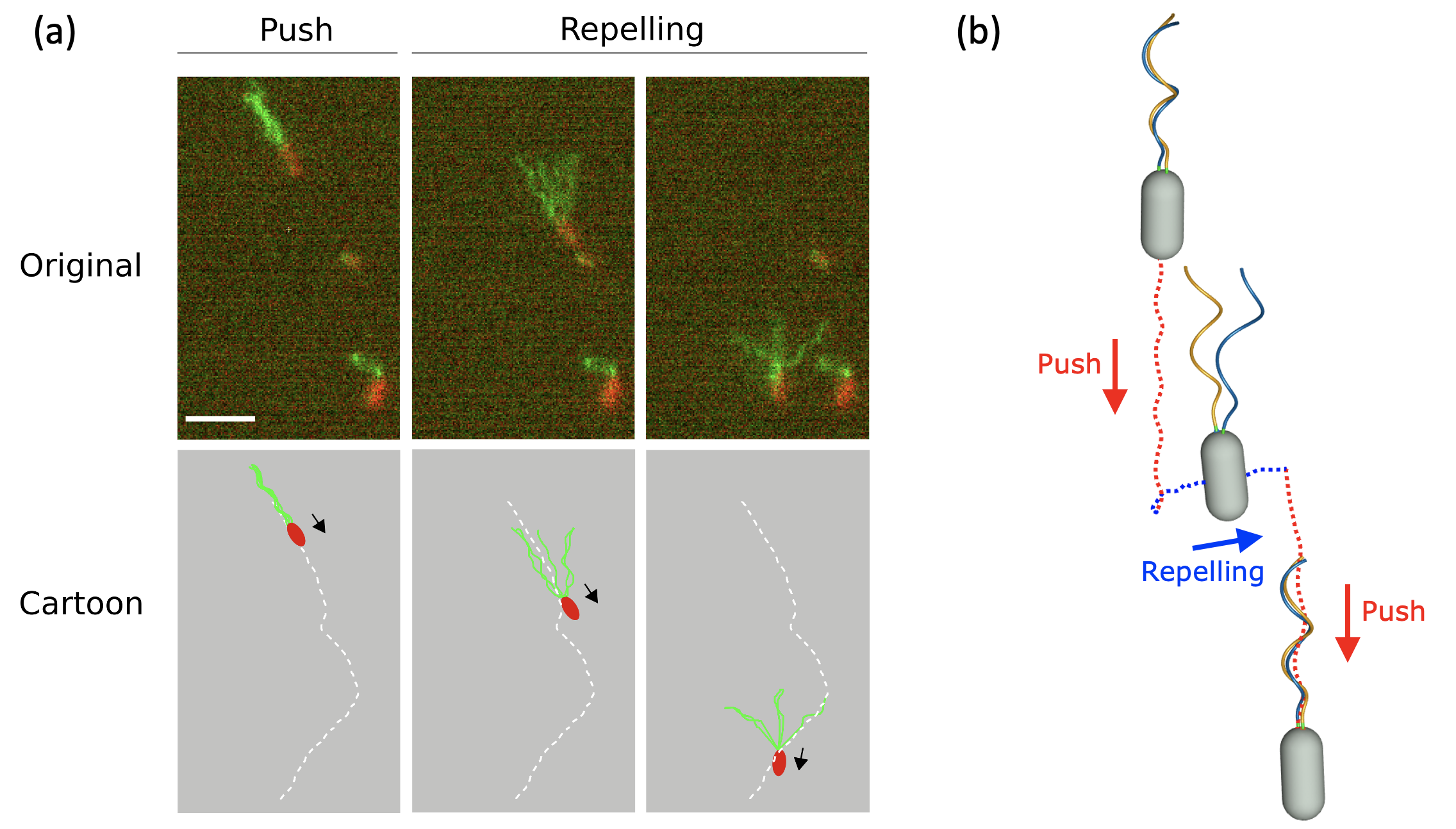}
	\caption{
 Experimental evidence resembling the repelling mode (a)  and snapshots taken from a corresponding numerical simulation (b).
(a) Time-lapse sequence of dual-color fluorescence images (top) and corresponding cartoon representations (bottom) of a {\it P.~putida} cell undergoing actively driven decomposition of the flagellar bundle and spreading apart of the individual filaments, interrupting the smooth swimming motion of the bacterium (Multimedia available online). The cell body is displayed in red and the flagellar bundle in green; scale bar 5$\mu$m.  (b) A similar sequence of push, repelling, and push modes is visualized from a numerical simulation (Multimedia available online).}
\label{fig:push_repel_push}
\end{figure*}

Jamming is a distinct mode that occurs when both motors turn CW with small magnitudes, causing 
the flagella to become tangled near the motors. Consequently, individual flagella interlock and cannot be continuously
rotated, unlike in other swimming modes. 
This entanglement phenomenon has been reported in both macroscopic experiments and theoretical modeling\cite{macnab1977bacterial,macnab1977normal,kim2003macroscopic,tuatulea2020geometrical}.
Furthermore, our simulations present diverse configurations of unbundled flagella under conditions that there is a significant difference 
 in motor torques applied to two flagella or when the motors exhibit opposite rotational directions.
From these configurations, the repelling mode is most reminiscent to a flagellar maneuver regularly observed in fluorescence recordings of swimming {\it P.~putida} cells.
Here, it can be frequently seen how the flagellar bundle is actively driven apart, bringing the smooth swimming locomotion to an abrupt halt, for an example, see Fig.~\ref{fig:push_repel_push}(a) (Multimedia available online). A numerical simulation of a cell that swims in push mode, interrupted by a short episode of repelling, where both motors turn in opposite direction, closely resembles these experimental observations, see Fig.~\ref{fig:push_repel_push}(b) (Multimedia available online).
Similar experimental observations have been reported previously\cite{hintsche2017polar}, and were associated with stop events that are known from early recordings of cell body trajectories of {\it P.~putida} cells\cite{theves_bacterial_2013}.
Additionally, our numerical simulations showed cylindrical enveloping as well as a semi-wrapping mode in which a single flagellum becomes unwrapped from a wrapped bundle when the rotational motion of the bundle is interrupted 
during the wrapping mode, see Fig. \ref{fig:diff_torque}(i).

In lophotrichous bacteria, push, pull, and wrapping swimming modes, orchestrated by bundle rotation and synchronous 
motor reversals, are predominant. Each mode mimics a single helix's behavior, with CCW or CW bundle rotation 
inducing wave propagation away from or toward the motors in the surrounding fluid, respectively.  
However, interflagellar dynamics within a bundle present complexities. 
Our simulations illustrate that wave propagation towards the motors in pull mode induces
a greater degree of flagellar intertwist than push mode, in which  
wave propagation towards the distal end of the flagella assists in limiting excessive intertwist. Additionally, the radius and pitch of the bundle helix in pull mode exceed intrinsic helical properties, resulting in a further reduction in the degree of flagellar torsion at the individual filament level.

It has been reported that flexural rigidity of flagella plays an important role in the flagellar dynamics\cite{lee2023bio, park2022modeling, brown2012flagellar, kuhn2018spatial,  shum2012}. Our simulations also emphasize the significance of elastic properties of flagella. 
When all  motors turn CW, a minimal torque is required to prevent the flagella from jamming at a fixed bending modulus of the filament. 
This threshold increases as the hook becomes more flexible, while a lack of hook flexibility increases the likelihood of jamming.
When all motors turn CCW, the threshold of motor torque that separates overwhirling from push increases as the hook becomes rigid. 
However, the absence of the hook predominantly favors the push mode, except in extreme cases involving high motor torque and low bending modulus of the filament or low motor torque and high bending modulus of the filament.  
This suggests that lophotrichous bacteria maintain an optimal range of the hook bending modulus to minimize the possibility of bundling failure and maximize swimming performance. Given that bundling is required for transitioning to the wrapping mode in {\it P. putida}, which is a beneficial swimming strategy in its native habitats, the significance of a flexible hook in {\it P. putida} cannot be overstated.

Smooth swimming with a stable bundle relies on the coordinated behaviors of multiple flagellar motors, facilitated by the cell's chemosensory system. Our future work will involve integrating a chemosensory system into our model to regulate the speed and direction of flagellar motors\cite{alirezaeizanjani2020chemotaxis, pfeifer2022role,thormann2022wrapped, nava2020novel}. This will deepen our understanding of how bundle stability influences the chemotactic behaviors of lophotrichous bacteria, particularly regarding turning angles during mode transitions\cite{alirezaeizanjani2020chemotaxis}, an area that has received less attention compared to peritrichous bacteria. Furthermore, it was speculated that bundle formation in pull mode may become unstable under specific conditions, such as high fluid viscosity or proximity to a surface\cite{kim2003macroscopic, macnab1977bacterial, thormann2022wrapped}, which may alter the swimming statistics of a lophotrichous swimmer close to surfaces and under confinement\cite{theves_random_2015}. We plan to incorporate
these conditions into our future study to further explore this instability. 

\section*{Supplementary Material}
 
  See the supplementary material for a table for parameters (Table S1), experimental methods (Text S1), a figure (Fig. S1), and videos (Video S1-S7).

\begin{acknowledgments}
S.L. was supported
by NSF (DMS-1853591 \& CBET-2415406) and the Charles Phelps Taft
Research Center at University of Cincinnati, USA. W.L. was supported by the National Institute for Mathematical
Sciences Grant funded by the Korean government
(B24910000).
Y.K. was supported by National Research Foundation of Korea Grant funded by the Korean government 
(RS-2023-00247232).
J.P. was supported by Research \& Creative Projects Award at the State University of New York at New Paltz, USA.
V.P. and C.B. received support by the Deutsche Forschungsgemeinschaft (DFG), project ID 443369470 - BE 3978/13-1.
V.M. acknowledges support by the International Max Planck Research School on Multiscale Bio-Systems.
\end{acknowledgments}

\section*{COMPETING INTERESTS} 
The authors have no conflicts to disclose.

\section*{Author contributions}
S.L. and J.P. designed research; S.L. J.P., Y.K. and W.L. performed numerical research and analyzed data; 
C.B. designed experimental research; V.P and V.M. performed experiments and analyzed experimental data; 
S.L., J.P., Y.K., W.L.  and C.B.  wrote the paper.

\section*{DATA AVAILABILITY}
The data that support the findings of this study are available from the corresponding author upon reasonable request.

\section*{REFERENCES}

%

\end{document}